# The Divide-and-Conquer Framework:
# A Suitable Setting for the DDM of the Future

*Ismael Herrera-Revilla\*, Iván Contreras and Graciela S. Herrera*

Instituto de Geofísica, Universidad Nacional Autónoma de México (UNAM), Mexico City, Mexico

\*Corresponding author: iherrerarevilla@gmail.com

This paper was prompted by numerical experiments we performed, in which algorithms already available in the literature (DVS-BDDM) yielded accelerations (or *speedups*) many times larger (more than seventy in some examples already treated, but probably often much larger) than the number of processors used. Based on these outstanding results, here it is shown that believing in the standard *ideal speedup*, which is taken to be equal to the number of processors, has limited much the performance goal sought by research on domain decomposition methods (DDM) and has hindered much its development, thus far. Hence, an improved theory in which the *speedup goal* is based on the Divide and Conquer algorithmic paradigm, frequently considered as the *leitmotiv* of domain decomposition methods, is proposed as a suitable setting for the DDM of the future.

## 1. Introduction

The present paper was prompted by some outstanding results that we recently obtained in a sequence of numerical and computational experiments applying some parallel algorithms already available in the literature (the DVS-BDDC [1-8]). They are outstanding, because they contradict the generally accepted belief that in parallel computation the acceleration, or *speedup*, cannot be greater than the processors number [9-22]. For example, in our numerical experiments using 400 processors in parallel we achieve a speedup of 29,278, which is 73.2 times greater than the maximum acceleration that such a belief permits.

In agreement with such a belief, the speedup goal sought in most, probably all, research that has been carried out in domain decomposition methods (DDM) up to now [9-20], is equal to the number of processors used. Since our results show that considerably larger speedups are feasible, the conclusion is drawn that the speedup goal sought so far is too modest and restrictive; hence, it should be replaced by a larger and more ambitious performance goal in future DDM research. To this end, we resort to the DIVIDE AND CONQUER STRATEGY, which for parallel processing of partial differential equations probably is the most basic algorithmic paradigm [23]. Furthermore, we formulate it in a manner that yields precise and clearly defined quantitative performance goals, to be called DC-goals, which are larger, yet realistic. The adequacy of the modified framework so obtained is verified by satisfactorily incorporating the outstanding results just mentioned in it.

The paper is organized as follows, Section 2 presents some background material on the Derived-Vector Space (DVS) approach to DDM, and DVS-BDDC [1-8]. The outstanding performance results that prompted this article are introduced and explained in Section 3. An inconsistency of standard approaches to DDM that such results exhibit, is pointed out and discussed in Section 4, while Section 5 introduces some measures of performance whose conspicuous feature is that they are defined with respect to a performance goal.

The ideas and results contained in Sections 3 to 5, are then used in Sections 6 and 7, to show that both, the concept of *ideal parallel-performance* and the belief that the *ideal parallel speedup* is $p$, lack firm bases. The *"divide and conquer"* algorithmic paradigm ([23], p. *v*), -the *DC-paradigm*- is recalled and revised in Section 8, and a quantitative *DC-performance goal* adequate to be used in future DDM research, is derived from it. There, it is also shown that in the examples here discussed the latter *performance goal* is larger, by a big factor, than $p$; indeed, in the examples here treated, the *DC-speedup goal* is close to $p^2$ and the factor we are referring is close to $p = p^2/p$, which is large when the number of processors is large.

When the outstanding numerical and computational results that prompted this paper are incorporated in the *DC-framework* they look perfectly normal, as it is shown in Section 10, since their *DC-efficiencies*, for $p \neq 1$, range from $70.3\%$ to $20.0\%$. Sections 11 and 12 are devoted to exhibit the severe restrictions that believing in the relation: $S(p,n) \leq p$ has imposed on software developed under that assumption. Finally, Section 13 states the paper's conclusions.

## 2. Some background

I. Herrera, and some of his coworkers, have been working in domain decomposition methods (DDM) since 2002, when he organized and hosted the Fourteenth International Conference on Domain Decomposition Methods (DDM) [23].

In works before the present paper [1-8], they have indicated that the use of coarse meshes in which some of the nodes are shared by several subdomains is a serious handicap because it goes against the *'divide and conquer paradigm'*, and then the system matrix so obtained is not block-diagonal. To overcome such an inconvenience, they introduced: the *derived-nodes*, the *derived-vectors* and the *derived-vector space* (DVS), which altogether yielded the *derived-vector space framework* (*DVS-framework*). The main advantage of DVS formulations is that the system matrix is strictly block-diagonal, while those of standard approaches are not [2].

In the *DVS algorithmic framework,* the procedure is as follows: Firstly, the partial differential equation is discretized in a *non-overlapping* fashion. The concept of *non-overlapping discretization* is given in reference [2]. The most significant and conspicuous property of such a kind of discretizations is that they yield block-diagonal systems of equations directly because of the discretization process and such systems of equations are defined in non-overlapping systems of nodes. A *non-overlapping* system of nodes is one that belongs to one and only one subdomain of the rough mesh (Fig. 1)

There are four DVS-algorithms: DVS-FETI-DP, DVS-BDDC, DVS-PRIMAL and DVS-DUAL. The first two were obtained by mimicking the well-known FETI-DP and BDDC procedures in the *derived-vector space*. (see, [2] for further details), but the big and very significant difference is that such procedures are applied after the differential equations have been subjected to a *non-overlapping discretization*, so that the discrete system of linear equations we start with, is block-diagonal. The other two DVS-algorithms, the DVS-PRIMAL and DVS-DUAL, were produced by completion of the theoretical framework (again see, [2]). So far, only the DVS-BDDC algorithm has been numerically tested; in 2016, preliminary computational experiments were published, which proved that the DVS-BDDC was fully competitive with the top DDM algorithms that were available [1]. However, at that time we did not have yet produced the very outstanding results we are now reporting.

## 3. The outstanding results

More recently, in 2018, the authors have developed a more careful code of the DVS-BDDC algorithm and tested it through a set of numerical experiments, obtaining the outstanding results that are presented and discussed in this Section. They are objectively outstanding because, for example, when the number of processors used is 400 the acceleration produced is 73.2 times by 400; hence, in this application, the DVS-BDDC algorithm produces an acceleration 73.2 times larger than the largest possible according to standard theory.

More specifically, the computational experiments here reported, consisted in treating a well posed 2D problem for Laplace differential operator in the highly parallelized supercomputer "Miztli" of the National Autonomous University of Mexico (UNAM), using successively 1, 16, 25, 64, 256 and 400 processors. The notation used to report the numerical and computational results so obtained is given next:

$$\begin{aligned} &\textit{Number of Processors} \to p \\ &\textit{Size of the Problem} \to n \\ &\textit{Execution Time} \to T(p,n) \\ &\textit{Speedup} \to S(p,n) \end{aligned} \quad (3.1)$$

Here, the size of the problem is equal to number of degrees of freedom, which in turn is equal to the number of nodes of the fine mesh. In general, the "execution time" and "speedup" are functions of the pair $(p,n)$. In the set of experiments here reported the size of the problem is expressed in terms of the number of degrees of freedom, which is taken to be the number of nodes of the fine mesh, which is kept fixed and equal to $10^6$; i.e., $n = 10^6$ throughout the set of numerical experiments.

The very impressive results of the numerical experiments are given in Table 1 (everywhere in this paper times are given in seconds), where the fifth column gives the speedup as a multiple of $p$, the number of processors, which in standard theory of domain decomposition methods is thought to be an unsurmountable speedup. However, in the set of experiments we are reporting, the *speedup* is much greater than the standard theory foresees, if $p \neq 1$; even more, it is greater than such an upper bound, by a large factor: 10.28, 15.02, 28.58, 57.18 and 73.2, when the number of processors is 16, 25, 64, 256 and 400, respectively. Observe that the factor increases with the number of processors, which is an enhancing feature. The last column is only included here, for later use.

| $p$ | $n$ | $T(p,n)$ | $S(p,n)$ | $S(p,n)$ in terms of $p$ | $p/S(p,n)$ |
|---|---|---|---|---|---|
| 1 | $10^6$ | 29,278 | 1 | 1p | 1 |
| 16 | $10^6$ | 178 | 164.5 | 10.28p | .097 |
| 25 | $10^6$ | 78 | 375.4 | 15.02p | .067 |
| 64 | $10^6$ | 16 | 1,829 | 28.58p | .035 |
| 256 | $10^6$ | 2 | 14,639 | 57.18p | .017 |
| 400 | $10^6$ | 1 | 29,278 | 73.20p | .014 |

Table 1. Results of computational experiments

## 4. An inconsistency of standard DDM

The standard definition of efficiency is

$$E_S(p,n) \equiv p^{-1} S(p,n) \qquad (4.1)$$

and it is usually expressed in percentage. The sub-index $S$ used here, comes from standard and it is used for clarity, since alternative definitions will be introduced later.

The Table 2 that follows has been derived from Table 1, by expressing its last column in terms of standard efficiency, $E_S(p,n)$. By inspection of Table 2, where percentages much greater than 100% such as 1,028, 1,502, 2,858, 5,718 and 7,320 occur, it is seen that the standard efficiency is not adequate for expressing the outstanding results of the numerical experiments we are reporting, because efficiencies far beyond 100% occur.

| $p$ | $n$ | $T(p,n)$ | $S(p,n)$ | $E_S(p,n)$ in percentage |
|---|---|---|---|---|
| 1 | $10^6$ | 29,278 | | 100% |
| 16 | $10^6$ | 178 | 164.5 | 1,028% |
| 25 | $10^6$ | 78 | 375.4 | 1,502% |
| 64 | $10^6$ | 16 | 1,829 | 2,858% |
| 256 | $10^6$ | 2 | 14,639 | 5,718% |
| 400 | $10^6$ | 1 | 29,278 | 7,320% |

Table 2. Using standard efficiency for expressing outstanding results

## 5. Revisiting the measures of performance

In this Section we define some measures of parallel-software performance that will be used in the sequel. As usual, such measures will be based on the execution time that is required for completing a task; the shorter the better. According to Eq.(3.1), the notation $T(p,n)$ means the execution time when the number of processors is $p$; in particular, $T(1,n)$ is the execution time when only one processor is applied.

For the sake of clarity, we recall the *speedup* (or, *acceleration*) definition:

$$S(p,n) \equiv \frac{T(1,n)}{T(p,n)} \tag{5.1}$$

The main objective in using a parallel computer is to get a simulation to finish faster than it would in one processor. Furthermore, let us take the position of a software designer who intends to construct software that performs well; so, he defines a *performance goal* he intends to achieve. The following two procedures for specifying such a goal will be considered; fixing the *execution-time goal*, $T_G(p,n)$, or fixing the *speedup goal*, $S_G(p,n)$. Assume either one of them have been specified, then the *relative efficiency* (relative to a goal performance) is defined by

$$E_G(p,n) \equiv \frac{S(p,n)}{S_G(p,n)} \tag{5.2}$$

when $S_G(p,n)$ is given, or

$$E_G(p,n) \equiv \frac{T_G(p,n)}{T(p,n)} \tag{5.3}$$

when $T_G(p,n)$ is given.

These two manners of defining *relative efficiency* are equivalent, if and only if:

$$T_G(p,n) S_G(p,n) = T(1,n) \tag{5.4}$$

Hence

$$S_G(p,n) = \frac{T(1,n)}{T_G(p,n)} \text{ and } T_G(p,n) = \frac{T(1,n)}{S_G(p,n)} \tag{5.5}$$

The first one of these equalities can be used to obtain $S_G(p,n)$ when $T_G(p,n)$ is given, and the second one, conversely.

According to Eq.(5.2),

$$E_G(p,n) = 1 \Leftrightarrow S(p,n) = S_G(p,n) \tag{5.6}$$

Here the symbol $\Leftrightarrow$ stands for the logical equivalence; i.e., if and only if. Actually, when we choose a goal we do not know if it is achievable, but the initial state satisfies $S(p,n) < S_G(p,n)$ since $S_G(p,n)$ is a desirable state. Hence, at the beginning $1 - E(p,n) > 0$ and this quantity may be taken as a distance to the goal. However, it can also happen that our developments lead to a speedup $S(p,n) > S_G(p,n)$, since generally we do not know beforehand if the speedup $S_G(p,n)$ is an upper bound of those possible. When that happens, $E(p,n) > 1$.

Conversely, a corresponding argument can be made if the execution time and Eq.(5.3) are used to define the parallel efficiency. The main difference is that, in such a case, $T(p,n) > T_G(p,n)$ at the beginning and $T(p,n) < T_G(p,n)$ is an indication that the goal has been exceeded.

## 6. The concept of "ideal parallel speedup"

In the literature on scientific parallel computing and on domain decomposition methods for the numerical solution for partial differential equations, the notion of *"ideal parallel* *speedup"* is used when defining absolute efficiency. However, its definition lacks precision. When $S_A(p,n)$ is the *ideal parallel speedup*, the relation

$$S(p,n) \leq S_A(p,n) \tag{6.1}$$

holds whenever $S(p,n)$ is the acceleration obtained in a parallel computation. If we try to make this notion rigorous, we could say that $S_A(p,n)$ is the *supremum*, but what is never made clear is: of what set $S_A(p,n)$ is the supremum.

Even so, when $S_A(p,n)$ is the *ideal parallel speedup*, the *absolute parallel efficiency* is defined to be

$$E_A(p,n) \equiv \frac{S(p,n)}{S_A(p,n)} \tag{6.2}$$

Thereby, we mention that the subscript $A$ above, comes from *Absolute*.

However, if we do not know for sure that Eq.(6.1) holds whenever $S(p,n)$ is the acceleration obtained in a parallel computation, this is a risky definition. Indeed, if that is the case and there is an execution for which

$$S(p,n) > S_A(p,n) \tag{6.3}$$

Then, we would claim that $S(p,n)$ is not achievable and we would be satisfied with an acceleration that is close to $S_A(p,n)$, even if $S_A(p,n)$ is much smaller than $S(p,n)$.

## 7. The international DDM research goal

In the light of the outstanding results we are reporting, it seems that is what has happened in the case of the international DDM research.

It is generally thought that Eq.(6.1) holds with $S_A(p,n) \equiv p$; i.e.,

$$S(p,n) \leq p \tag{7.1}$$

Hence, the standard definition of efficiency of Eq.(4.1):

$$E_S(p,n) \equiv p^{-1} S(p,n) \tag{7.2}$$

Comparing this equation with Eq.(5.2) it is seen that Eq.(7.2) implies that the speedup goal, sought by DDM research worldwide is:

$$S_S(p,n) \equiv p \qquad (7.3)$$

Here, we have written $S_S(p,n)$ for the *speedup goal* of standard DDM research.

## 8. The relative DVS efficiency of standard approaches

In this Section we make a simple exercise in which we compute the *relative efficiency* of standard approaches when the *goal speedup* is that achieved by the DVS-BDDC algorithm in the numerical experiments here reported. The notation here adopted for such a *relative efficiency* is $E_{DVS}^S$.

Applying the definition of Eq.(5.2), we get

$$E_{DVS}^S(p,n) = \frac{S(p,n)}{S_G(p,n)} \leq \frac{p}{S_{DVS}(p,n)} \qquad (8.1)$$

Inspecting the results of our numerical experiments reported in the last column of Table 1, in view of Eq.(7.1), it is seen that the *relative efficiency* of standard approaches relative to the performance of DVS-BDDC is only 9.7%, 6.7%, 3.5%, 1.7% and 1.4%, of what is obtained with DVS-BDDC in these experiments. Hence, our conclusion of this Section is that the speedups goals sought by DDM research worldwide up to now, are too small and should be revised.

## 9. The Speedup Goal of the Divide and Conquer Framework

As a starting point for that purpose, we recall the **divide and conquer** algorithmic paradigm ([23], p.v), which is frequently considered as the *leitmotiv* of domain decomposition methods [21].

The **divide and conquer strategy** (*DC-strategy*) consists in dividing the domain of definition of the scientific or engineering model into small pieces and then send each one of them to different processors. If $p$ is the number of subdomains of the domain decomposition, the size of each piece is approximately equal to $n/p$; hence, smaller than $n$ when $p > 1$ and much smaller than $n$, when $p$ is large.

This is the procedure for reducing the size of the problems treated by the different processors when the *DC-strategy* is applied. Of course, for the **divide and conquer strategy** being effective it is necessary and sufficient that each one of the *local problems* be independent of all others. Such a condition (each *local problem* being independent of all others) is seldom fulfilled in practice, and it will be referred to as the *DC-paradigm*. Adopting the *DC-paradigm* as a guide in the development of software implies to strive to construct algorithms in which the local problems are as independent from each other as possible. Thereby, we mention that the *derived-vector space* framework, which in the outstanding numerical experiments here reported has been so effective, was developed following the *DC-paradigm*.

Since the approximate size of each *local problem* is $n/p$, when all them are independent, $T(1, n/p)$ would be the approximate *execution-time* for each one of them, which when the computation is carried out in parallel is also the global *execution-time*. Therefore, in the *DC-framework* we define the *execution-time goal* (*DC-execution-time goal*), to be denoted by $T_{DC}(p,n)$, as:

$$T_{DC}(p,n) \equiv T(1, n/p) \tag{9.1}$$

Correspondingly, the *speedup goal* for the DC-approach is defined to be

$$S_{DC}(p,n) \equiv \frac{T(1,n)}{T(1,n/p)} \tag{9.2}$$

and the *DC-efficiency* is given by

$$E_{DC}(p,n) \equiv \frac{S(p,n)}{S_{DC}(p,n)} = \frac{T(1,n/p)}{T(p,n)} \tag{9.3}$$

In Table 3, to illustrate the *Divide and Conquer* concepts, they have been computed in the conditions of the numerical experiments that prompted this paper. The first and second columns (counted from left to right) contain the number of processors and the degrees of freedom of the local problems, respectively. The third column yields the DC-execution time goals of the local problems, which were obtained through numerical experiments; for each $p$ only one of the local problems was solved numerically (and only one of the processors was used). Once $T_{DC}(p,n)$ was known, $S_{DC}(p,n)$ was computed applying straightforward formulas. The local solvers used in our numerical experiments were banded *LU* decompositions, whose *algorithmic complexity* turned out to be $p^2$ and is given in the fifth column. An interesting fact, in the numerical experiments here reported, is that the *algorithmic complexity* approximates $S_{DC}(p,n)$, and the last column of Table 3 gives the corresponding relative errors in percentage associated with such an approximation. For the purpose we have in mind, such errors are admissible.

| $p$ | $n/p$ | $T_{DC}(p,n)$ | $S_{DC}(p,n)$ | $p^2$ | $\{p^2 - S_{DC}(p,n)\}/p^2$ |
|---|---|---|---|---|---|
| 1 | $10^6$ | 29,278 | 1 | 1 | 0% |
| *16* | 62,500 | 125.15 | 233.9 | 256 | 8.6% |
| 25 | 40,000 | 51.45 | 596.1 | 625 | 4.6% |
| 64 | 15,625 | 7.90 | 3,706 | 4,096 | 9.5% |
| 256 | 4,096 | 0.55 | 53,233 | 65,536 | 18.8% |
| 400 | 2,500 | 0.2 | 146,390 | 160,000 | 8.5% |

TABLE 3 The *DC-execution-time goal* and the *DC-speedup goal*

## 10. Incorporating the outstanding results in the DC-framework

In this Section the results of our numerical and computational experiments contained in Table 1, are incorporated in the DC-framework. Table 4 that was so built follows. The seventh column of Table 4 gives the efficiency relative to the Divide and Conquer performance goal, while the last column gives the same efficiency when it is estimated approximating $S_{DC}(p,n)$ by $p^2$.

| $p$ | $p^2$ | $T(p,n)$ | $S(p,n)$ | $T_{DC}(p,n)$ | $S_{DC}(p,n)$ | $E_{DC}(p,n) \equiv$ $T_{DC}(p,n)/T(p,n)$ $= S(p,n)/S_{DC}(p,n)$ | $S(p,n)/p^2$ |
|---|---|---|---|---|---|---|---|
| 1 | 1 | 29,278 | 1 | 29,278 | 1 | 100% | 100% |
| 16 | 256 | 178 | 164.5 | 125.15 | 233.9 | 70.3% | 64.3% |
| 25 | 625 | 78 | 375.4 | 51.45 | 596.1 | 63.0% | 60.1% |
| 64 | 4,096 | 16 | 1,829 | 7.90 | 3,706 | 49.4% | 44.7% |
| 256 | 65,536 | 2 | 14,639 | 0.55 | 53,233 | 27.5% | 22.3% |
| 400 | 160,000 | 1 | 29,278 | 0.2 | 146,390 | 20.0% | 18.3% |

Table 4. The outstanding results in the DC-framework

By inspection of this table, it is seen that the *outstanding results* that prompted this paper look perfectly normal when they are displayed in the *DC-framework*. This shows that the *DC-framework* is adequate for accommodating the outstanding numerical and computational results that we have obtained using the *DVS-BDDC* algorithm.

## 11. Restrictions on parallel performance imposed by the standard framework

Assuming $S(p,n) \leq p \equiv S_S(p,n)$ is limitative and this Section together with the next one we explore more thoroughly the restrictions on parallel performance that such an assumption imposes.

To start with, the *standard speedup goal*, $p$, and the *DC-speedup goal*, $S_{DC}(p,n)$, corresponding to the set of experiments we have been discussing, are compared. Their ratios are shown Table 5, where the values of $S_{DC}(p,n)$ are taken from Table 4.

| $p$ | $S_S(p,n)$ | $S_{DC}(p,n)$ | $S_{DC}(p,n)/S_S(p,n)$ $= S_{DC}(p,n)/p$ | $S_S(p,n)/S_{DC}(p,n)$ $= p/S_{DC}(p,n)$ |
|---|---|---|---|---|
| 1 | 1 | 1 | 1 | 1 |
| 16 | 16 | 233.9 | 14.6 | .0685 |
| 25 | 25 | 596.1 | 23.8 | .0420 |
| 64 | 64 | 3,706 | 57.9 | .0173 |
| 256 | 256 | 53,233 | 207.9 | .0048 |
| 400 | 400 | 146,390 | 365.0 | .0027 |

Table 5. Comparison of *speedup goals*

By inspection of Table 5, it is seen that the *standard goal-speedups* are much smaller than the *goal-DC-speedups*, and probably too conservative and restrictive.

Table 6, which follows, shows the bounds of performance for any software that satisfies the

| $p$ | $T_S(p,n) \geq T(1,n)/p$ | $S_S(p,n) \leq p$ | $T_{DC}(p,n)$ | $S_{DC}(p,n)$ | $E_{DC}^S(p,n) \leq p/S_{DC}(p,n)$ |
|---|---|---|---|---|---|
| 1 | – | – | 29,278 | 1 | 100% |
| 16 | $T_S(16,10^6) \geq 1,830$ | $S_S(16,10^6) \leq 16$ | 125.15 | 233.9 | $E_{DC}^S(16,10^6) \leq 6.85\%$ |
| 25 | $T_S(25,10^6) \geq 1,171.1$ | $S_S(25,10^6) \leq 25$ | 51.45 | 596.1 | $E_{DC}^S(25,10^6) \leq 4.20\%$ |
| 64 | $T_S(64,10^6) \geq 457.5$ | $S_S(64,10^6) \leq 64$ | 7.90 | 3,706 | $E_{DC}^S(64,10^6) \leq 1.73\%$ |
| 256 | $T_S(256,10^6) \geq 114.40$ | $S_S(256,n) \leq 256$ | 0.55 | 53,233 | $E_{DC}^S(256,10^6) \leq 0.48\%$ |
| 400 | $T_S(400,10^6) \geq 73.20$ | $S_S(400,n) \leq 400$ | 0.20 | 146,390 | $E_{DC}^S(400,10^6) \leq 0.27\%$ |

Table 6. Restrictions of performance for standard software

restriction $S(p,n) \leq p$. The last column of this table shows such an assumption limits severely the *DC-efficiency* that one can hope for, when any of the standard methods is applied, including BDDC and FETI-DP [22].

**12. Additional comparisons**

To have a clearer appreciation of the relevance of the limitations imposed by the standard framework, which have been established in Section 9, a direct comparison with the results obtained using the DVS-BDDC, which are given in Table 3, can help. Such a comparison is highlighted in Table 7.

| Method\p | 16 | 25 | 64 | 256 | 400 |
|---|---|---|---|---|---|
| DVS-BDDC DC-efficiencies | 70.3% | 63.0% | 49.4% | 27.5% | 20.0% |
| DC-efficiencies: Bounds for standard | 6.85% | 4.20% | 1.73% | 0.48% | 0.27% |

Table 7. Direct comparison of *DC-efficiencies* of standard and DVS-BDDC software

In summary, for all the numerical and computational experiments here discussed, the efficiency one can hope to obtain using standard software is only a small fraction of that, which is obtained when the DVS-BDDC algorithm is applied.

From all the above discussion, we draw the conclusion that adopting the definition $S_S(p,n) \equiv p$, as is usually done in domain decomposition methods, is too conservative and hinders drastically the performance of methods developed within such a framework.

**13. Conclusions**

This paper communicates the outstanding results of numerical experiments in which the DVS-BDDC algorithm [2] yields speedups exceeding the number of processors multiplied by a large factor; 73.2 is the largest obtained in our experiments. From the analysis here presented the following conclusions are drawn:

1. The belief that the speedup (or, acceleration) is always less or equal to $p$ (the number of processors) is incorrect. Accelerations much larger than $p$ are not only feasible, but have been achieved using the DVS-BDDC algorithm;

2. The *performance goal* that research on DDM has intended up to now, besides being too small, has been very restrictive for the software developed in that framework; and

3. The *Divide and Conquer framework*, here introduced, is more adequate for accommodating results such as those that have been obtained using the *DVS-BDDC* algorithm.

Based on these conclusions it recommended that the *Divide and Conquer framework* be adopted in future research on the applications of parallel computation to the solution of partial differential equations. Then, the *performance goal* is defined in terms of the *execution time goal*, as

$$T_{DC}(p,n) \equiv T(1, n/p) \tag{13.1}$$

Or, the *speedup goal*,

$$S_{DC}(p,n) \equiv \frac{T(1,n)}{T(1,n/p)} \qquad (13.2)$$

Or, the *divide and conquer efficiency*:

$$E_{DC}(p,n) = \frac{T(1,n/p)}{T(p,n)} = \frac{S(p,n)}{S_{DC}(p,n)} \qquad (13.3)$$

**Acknowledgment**

We want to thank DGTIC for its support and computational resources assigned to this research in the cluster Miztli, which is the supercomputer of the National Autonomous University of Mexico (UNAM) under project LANDCAD-UNAM-DGTIC-065.